\documentclass{article}

\usepackage{amsmath}
\usepackage{PRIMEarxiv}
\usepackage{subcaption}
\usepackage[utf8]{inputenc} % allow utf-8 input
\usepackage[T1]{fontenc}    % use 8-bit T1 fonts
\usepackage{hyperref}       % hyperlinks
\usepackage{url}            % simple URL typesetting
\usepackage{booktabs}       % professional-quality tables
\usepackage{amsfonts}       % blackboard math symbols
\usepackage{nicefrac}       % compact symbols for 1/2, etc.
\usepackage{microtype}      % microtypography
\usepackage{lipsum}
\usepackage{authblk}
\usepackage{graphicx}
\graphicspath{{media/}}     % organize your images and other figures under media/ folder
  
\title{Combining Entangled and Non-Entangled based Quantum Key Distribution Protocol with GHZ state}

\author[$a$]{Arman Sykot}
\author[$b$, $c$]{Mohammad Hasibur Rahman}
\author[$d$, $c$]{Rifat Tasnim Anannya}
\author[$e$, $c$]{Khan Shariya Hasan Upoma}
\author[$a$,$\dagger$]{M.R.C. Mahdy}

\affil[$a$]{Department of Electrical and Computer Engineering, North South University, Dhaka, Bangladesh}
\affil[$b$]{Department of Computer Science and Engineering, University of Texas at Arlington 914 Greek Row Dr, Arlington, TX-76013, USA}
\affil[$c$]{Mahdy Research Academy, Dhaka, Bangladesh}
\affil[$d$]{Department of Computer Science, American International University Bangladesh, Dhaka, Bangladesh}
\affil[$e$]{Department of Electrical and Electronics Engineering, Sonargaon University, Dhaka, Bangladesh}
\affil[$\dagger$]{Corresponding author: M.R.C. Mahdy, mahdy.chowdhury@northsouth.edu}

\begin{document}

\maketitle

\begin{abstract}
This paper presents a novel hybrid Quantum Key Distribution (QKD) protocol that combines entanglement-based and non-entanglement based approaches to optimize security and the number of generated keys. We introduce a dynamic system that integrates a three-particle GHZ-state method with the two-state B92 protocol, using a quantum superposition state to probabilistically switch between them.
The GHZ-state component leverages strong three-particle entanglement correlations for enhanced security, while the B92 component offers simplicity and potentially higher key generation rates. Implemented and simulated using Qiskit, our approach demonstrates higher number of generated keys compared to standalone protocols while maintaining robust security.
We present a comprehensive analysis of the protocol's security properties and performance characteristics. The results show that this combined method effectively balances the trade-offs inherent in QKD systems, offering a flexible framework adaptable to varying channel conditions and security requirements.
This research contributes to ongoing efforts to make QKD more practical and efficient, potentially advancing the development of large-scale, secure quantum networks.
\end{abstract}

% keywords can be removed

\section{Introduction}
Quantum Key Distribution (QKD) \cite{Tomamichel2017largelyself} has emerged as a promising technology for secure communication in the quantum era. By leveraging the principles of quantum mechanics, QKD offers the potential for information-theoretic security, a feat unattainable with classical cryptographic methods. As we stand on the brink of a new technological paradigm, the importance of robust and efficient QKD protocols cannot be overstated.

Traditional QKD protocols can be broadly categorized into two types: those based on entanglement, such as the E91 \cite{ekert1991quantum} protocol, and those that do not require entanglement, like the BB84 \cite{bennett2014quantum} or B92 \cite{bennett1992quantum} protocols. Each approach has its strengths and limitations. Entanglement-based protocols offer stronger security guarantees through quantum correlations but can be more challenging to implement and may have lower key generation rates. Non-entanglement based protocols, on the other hand, can be simpler to implement and potentially faster, but may be more vulnerable to certain types of attacks.

In this paper, we propose a novel approach that combines the strengths of both entangled and non-entangled QKD protocols. Specifically, we introduce a hybrid protocol that integrates a GHZ-state \cite{greenberger1989going} based method (Entanglement-based protocol) with the B92 protocol (Non-entanglement-based protocol). The GHZ-state component leverages the strong correlations of three-particle entanglement, providing enhanced security, while the B92 component offers simplicity and potentially higher number of generated keys under certain conditions.

The primary goal of this combined approach is to achieve a higher overall key generation rate while maintaining the high level of security afforded by entanglement-based methods. By dynamically switching between the two protocols based on a quantum superposition state, we aim to optimize the trade-off between security and efficiency.

Our work builds upon the foundational research in quantum key distribution, including the seminal work of Bennett on B92, and the three-particle entanglement studies of Greenberger, Horne, and Zeilinger. We extend these ideas by proposing a unified framework that can adapt to varying channel conditions and security requirements.

In the following sections, we will provide a detailed description of our combined protocol, analyze its security properties, and present simulation results using the Qiskit \cite{QiskitCommunity2017} quantum computing framework. We will demonstrate how this hybrid approach can potentially outperform single-protocol implementations in terms of key generation rate while maintaining robust security guarantees.

Through this research, we aim to contribute to the ongoing efforts to make QKD more practical and efficient, bringing us closer to the realization of large-scale, secure quantum networks.

\section{Background Studies}
Quantum Key Distribution, known as QKD, is based on the quantum mechanics that aims to create secure cryptographic keys between authorized parties \cite{scarani2009security}. The paper provides an in-depth study of entanglement \& non-entanglement-based method within QKD. By exploring a thorough analysis of recent advancements, we highlight the significance of integrating these two methods. 
In QKD, two authorized entities represented as Alice \& bob, attempt to establish a secure communication over a distance using two channels.  A quantum channel, for the transmission of quantum signals; and a classical channel, to exchange classical messages. And the beauty of QKD is in the end, either they indeed successfully generate a secret key, or due to the possible tapping of an eavesdropper, typically referred as Eve, they abort the process. Furthermore, if key generate then according to Heisenberg’s uncertainty principle, these keys will remain secret from Eve.

Since the beginning of QKD in the 1980s, it has evolved as an entanglement-based and non-entanglement-based protocol, by offering distinct advantages in security, efficiency, and especially key distribution rates \cite{einstein1935can}. Early protocols like BB84 and later developments such as Ekert’s E91 protocol considered as the base of modern quantum cryptography. Greenberger-Horne-Zeilinger (GHZ) states are incorporated later, that improves multi-party quantum communication substantially. 

In 1984, BB84, the earliest QKD protocol, was proposed by Bennett and Brassard \cite{bennett1984proceedings}. Among four possible polarization states and by using random basis choice, key will generate. Based on the Heisenberg Uncertainty Principle \cite{phillips1992heisenberg} and the no-cloning theorem \cite{koashi1998no}, security of BB84 is ensured. The presence of Eve is detected for any attempt that will disturb the quantum states. While BB84 does not rely on the spookiest feature of QKD which is quantum entanglement but due to its simplified implementation, it has become one of the most widely implemented protocols in quantum communication systems \cite{Horodecki_2009}. However, a theoretical maximum efficiency of 50\% can be achieved in an ideal situation, if 50\% of the transmitted qubits are used for the final key generation. It also further reduced by error correction and privacy amplification. While BB84 uses four polarization states, B92 protocol simplifies it by utilizing only two non-orthogonal quantum states in 1992 \cite{bennett1992quantum}. This evolution led to higher key generation rates while also make it more prone to certain vulnerabilities, making it less secure than BB84. Notwithstanding these drawbacks, B92 is still a useful protocol where system performance takes precedence above absolute security. A variant of the BB84 protocol has been presented named SARG04 \cite{PhysRevLett.92.057901}, that differs primarily in its classical post-processing phase. It specially uses weak coherent light sources that is dedicatedly applied for Photon Number Splitting (PNS) attack. Compared to BB84, the key generation rate decreased in SARG04 and often achieve shorter secure distance. Another extended version of BB84 is known as six-state protocol which added a third orthogonal basis (circular polarization or Y basis) \cite{bruss1998optimal}. By using three bases instead of two, its enhanced resistance to eavesdropping although increased complexity. The study claims that by utilising solely one-way classical communication for error correction and privacy amplification, the Six-State protocol can withstand a bit error rate of up to 12.7\%. This is marginally more than BB84's 11\% error tolerance in comparable circumstances. However, this trade-off is frequently justified by the greater security, particularly in situations where efficiency is subordinated to security.

In XXI century, the subtlest feature was identified by EPR, known as quantum entanglement. The feature indicates that the presence of overall states in a combined system cannot be expressed as a product states of the individual subsystem \cite{von2013mathematische}. Simply, entanglement can be described as dependent connection between nonlocal measurement from where mutual information can be obtained. 

Artur Ekert proposed a groundbreaking Entanglement-based QKD protocols known as E91 protocol \cite{ekert1991quantum}. In contrast to BB84 and B92, E91 relies on quantum entanglement between pairs of particles shared between Alice and Bob. Ekert’s protocol is entrenched in Bell’s Theorem and the violation of Bell inequalities, which ensure that any attempt by an eavesdropper, Eve, to intercept the communication will disturb the entangled state in a detectable manner. In E91, a strong theoretical guarantee lies, even when separated by long distance, quantum mechanical nature of entanglement and Bell's inequality make it distinct in the field of QKD.

A noteworthy advancement is introduced in entanglement-based QKD protocols called Greenberger-Horne-Zeilinger (GHZ) states \cite{zukowski1998quest}. For securing multi party QKD, the entanglement has seen beyond typical bipartite entanglement. According to Guo et al. \cite{guo2010secure} the concept of nonlocality is demonstrated in the GHZ states. Even though the particles are separated by large distance still their measurement outcome is strongly correlated on the measurement performed on others. Either all the particles are in state “0” or “1” (2010). The integrity of the key preserved here from eavesdropper. In terms of Key generation efficiency, it achieve higher compared to bipartite protocol. Thus, considered a valuable tool for scaling QKD.

Bell’s Theorem and the violation of Bell inequalities act as a pivotal element in the security of entanglement-based QKD protocols \cite{bell1964physics}. In Bell inequalities, if the correlations between entangled particles violate the inequalities then it confirms the presence of entanglement while ensuring security of key distribution. E91 and GHZ-based QKD are grounded in this principle. An experiment was performed in the paper presented by Simon \cite{storz2023loophole} which demonstrates a loophole-free violation of Bell’s inequality using superconducting circuits. The violation of Bell's inequality was verified \cite{cafaro2024violationbellsinequalityclauserhorneshimonyholt} using the Clauser-Horne-Shimony-Holt (CHSH) inequality resulting a violation of Bell's inequality with an average S value of 2.0747. It is evident that any value greater than 2 confirms a violation of local realism. Nevertheless, the experiment suffered from photon loss, limiting the fidelity of the Bell state. In BBM92 \cite{bennett1992quantumbbm}, author has presented an entanglement-based version of the BB84 QKD protocol. In this protocol the randomness is inherent in the measurement of the entangled photon pairs. In addition, the key generation rate depends on several factors, including the quality of the entanglement and channel conditions. 

Even though both entangled and non-entangled QKD protocols have their own fortes but they also have limitations in terms of implementation, key generation rates, security. To address these trade-offs, we comprise a hybrid QKD protocols that allows high key generation rates, while still maintaining the security advantages of entanglement.

\section{Preliminary Concepts}
\subsection{Quantum States}
In quantum computing, a quantum state \cite{mcmahon2008quantum}, particularly a two-state system or qubit, can be expressed as a superposition of its basis states \( |0\rangle \) and \( |1\rangle \). An arbitrary qubit is described as \( |\psi\rangle = \alpha|0\rangle + \beta|1\rangle \), where \( |\alpha|^2 \) represents the probability of measuring the system in state \( |0\rangle \), and \( |\beta|^2 \) corresponds to the probability of finding it in state \( |1\rangle \).

\subsection{GHZ states}
The Greenberger-Horne-Zeilinger (GHZ) state \cite{greenberger1989going}, represented as 
\begin{equation*}
    |GHZ\rangle = \frac{1}{\sqrt{2}}(|000\rangle + |111\rangle)
\end{equation*}
for three qubits, plays a crucial role in quantum information theory and quantum key distribution. These states exhibit maximal entanglement, with a von Neumann entropy of 1 for any bipartition, making them ideal for multi-party quantum protocols. GHZ states provide a striking demonstration of quantum non-locality, violating local realism without requiring statistical arguments. 

\subsection{QKD B92}
The B92 \cite{bennett1992quantum} protocol uses two non-orthogonal quantum states for encoding bits. Alice generates a random bit string:
\begin{align*}
    |\psi_0\rangle &= |0\rangle & \text{(for bit 0)} \\
    |\psi_1\rangle &= \frac{|0\rangle + |1\rangle}{\sqrt{2}} & \text{(for bit 1)}
\end{align*}

Bob randomly chooses between two measurement basis:
\begin{itemize}
    \item \( B_0 = \{|\psi_1\rangle, |\psi_1^\perp\rangle\} \) to detect \( |\psi_0\rangle \),
    \item \( B_1 = \{|\psi_0\rangle, |\psi_0^\perp\rangle\} \) to detect \( |\psi_1\rangle \).
\end{itemize}
A bit is added to the key when Bob measuring $|1\rangle$ when expecting $|+\rangle$ indicates a '1' bit and when measuing $|-\rangle$ when expecting $|0\rangle$ indicates a '0' bit.

\section{Proposed Combined Protocol}
\begin{figure}[!h]
    \centering
    \includegraphics[width=.96\linewidth]{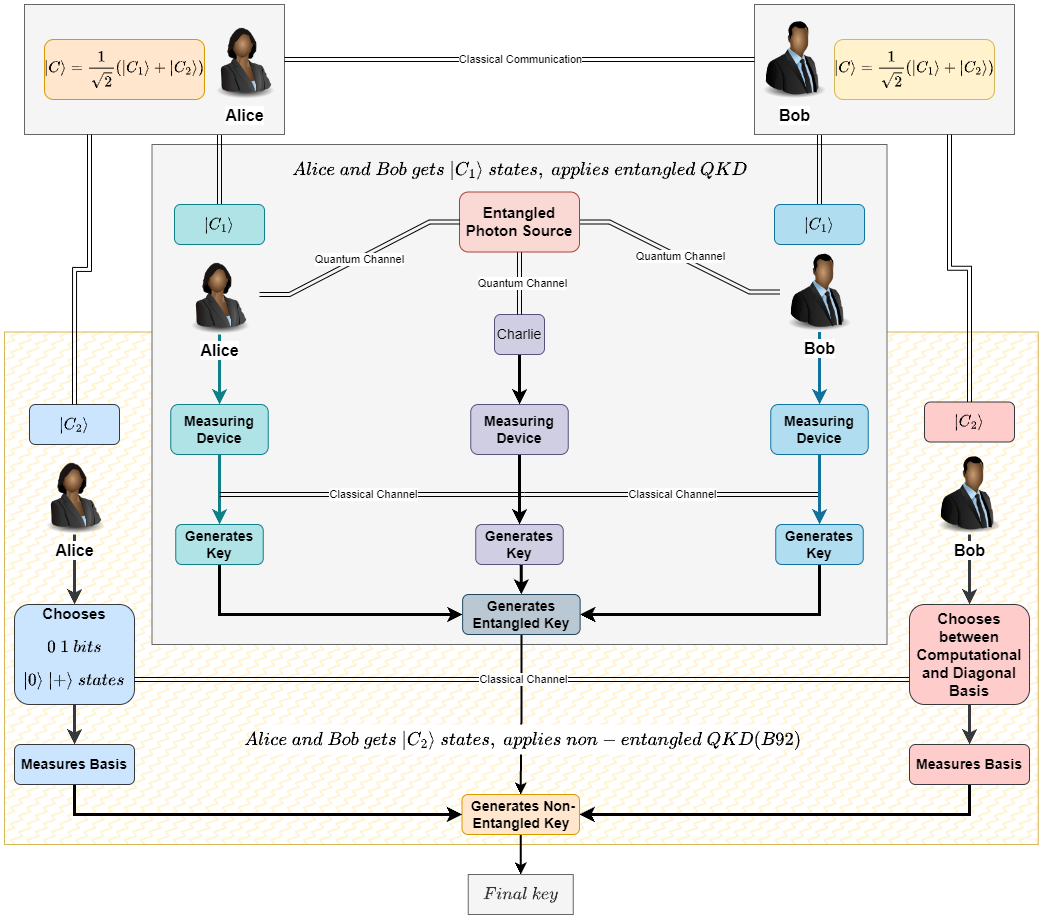}
    \caption{Schematic diagram of the combined entangled and non-entangled quantum key distribution (QKD) protocol. The protocol alternates between an entanglement-based QKD scheme involving Alice, Bob for |C1⟩ states, and a non-entangled QKD scheme (B92) between Alice and Bob for |C2⟩ states. The final key is generated by combining the outputs from both schemes for multiple rounds.}
    \label{fig:main figure}
\end{figure}

Our novel approach integrates the GHZ-state based protocol with the B92 protocol, using a quantum superposition state to dynamically select between them for multiple round to ensure combination of both protocol. This combined method aims to leverage the strengths of both protocols while mitigating their individual weaknesses. Figure: \ref{fig:main figure} shows the complete overview of our proposed architecture. 

\subsection{Superposition-based Selection Mechanism}
The protocol begins with a quantum coin flip \cite{molina2005experimental} to determine which sub-protocol to use for each round:
\begin{equation}
    |C\rangle = \frac{1}{\sqrt{2}} (|C_1\rangle + |C_2\rangle)
\end{equation}

If Alice and Bob's measurement falls in the same state, they choose the state to distribute key with the following protocol for multiple rounds, where $|C_1\rangle$ corresponds to the GHZ protocol and $|C_2\rangle$ corresponds to the B92 protocol. This qubit is measured in the computational basis, with the outcome determining the protocol choice:

\begin{itemize}
    \item If the result is $|C_1\rangle$, the GHZ protocol is used.
    \item If the result is $|C_2\rangle$, the B92 protocol is used.
\end{itemize}

The state $|C\rangle$ is normalized wave function. Since, $\langle C|C\rangle = 1$, where $\langle C| = \frac{1}{\sqrt{2}} (\langle C_1| + \langle C_2|)$.

So, probability of measuring $C_1$ is,
\begin{equation}\label{c1prob}
    P(C_1) = (\frac{1}{\sqrt{2}})^2 = \frac{1}{2}
\end{equation}

And probability of measurering $C_2$ is,
\begin{equation}\label{c2prob}
    P(C_2) = 1 - P(C_1) = \frac{1}{2}
\end{equation}
When measuring the state $|C\rangle$, there is a 50\% chance of obtaining $|C_1\rangle$ and a 50\% chance of obtaining $|C_2\rangle$. Hence, their is always equal change of getting GHZ protocol and B92 protocol through the superposition-based selection mechanism.
    
\subsection{GHZ-based QKD}
\textbf{State preparation:} The protocol involves three parties- Alice, Bob and Charlie. Charlie could be a device that belongs to Alice or Bob. The protocol begins withe the preparation of GHZ states, represented as: 
\begin{equation}
    |GHZ\rangle = \frac{1}{\sqrt{2}}(|0_A0_B0_C\rangle + |1_A1_B1_C\rangle)
\end{equation}

\textbf{Basis Selection:} Each party independently and randomly selects a measurement basis, either $\sigma_x = \begin{pmatrix}
  0 & 1\\ 
  1 & 0
\end{pmatrix}$ or $\sigma_y = \begin{pmatrix}
  0 & -i\\ 
  i &  0
\end{pmatrix}$, for each GHZ state. This random selection is crucial for the security of the protocol.

\textbf{Measurement:} Each party measures their qubit in their chosen basis and records the result. The parties announce their measurement basis (but not results) over a classical channel. 
When all parties measure in the $\sigma_y$ basis, their results should be correlated such that the product of their outcomes is always $-1$. They keep these results as key bits.
\begin{itemize}
    \item For $Y_AY_BY_C$ measurements: 
        \begin{equation}
            \langle \sigma_{yA} \sigma_{yB} \sigma_{yC} \rangle = -1
        \end{equation}
\end{itemize}

\textbf{Security check:} To detect eavesdropper, the parties can use a subset of their measurements (which was not used to generate key bits) to check the GHZ correlations \cite{greenberger1990bell}:
\begin{itemize}
    \item For $X_AX_BX_C$ measurements: 
        \begin{equation}\label{ghzcorr1}
        \langle \sigma_{xA} \sigma_{xB} \sigma_{xC} \rangle = +1     
        \end{equation}
    \item For $X_AY_BY_C$, $Y_AX_BY_C$, $Y_AY_BX_C$ measurements:
        \begin{equation}\label{ghzcorr2}
        \langle \sigma_{xA} \sigma_{yB} \sigma_{yC} \rangle = \langle \sigma_{yA} \sigma_{xB} \sigma_{yC} \rangle = \langle \sigma_{yA} \sigma_{yB} \sigma_{xC} \rangle = -1      
        \end{equation} 
\end{itemize}

These correlations cannot be explained by local hidden variable theories \cite{liu2021experimental}. Thus an eavesdropper cannot predict or copy these correlations without disturbing the state, which is detectable. 

\subsection{B92 Protocol}
\textbf{State preparation:} Alice generates a random bit string, for each bit:
\begin{itemize}
    \item If the bit is 0, Alice prepares the state in the $|0\rangle$ state.
    \item If the bit is 1, Alice prepares the state in the $|+\rangle$ state.
\end{itemize}

Alice sends these qubits to Bob through a quantum channel.

\textbf{Measurement:} Bob generates his own random bit string, for each received qubit:
\begin{itemize}
    \item If Bob's bit is 0, he measures in the $z$ (computational)-basis.
    \item If Bob's bit is 1, he measures in the $x$ (diagonal)-basis.
\end{itemize}
Bob announces which measurements gave him a conclusive result (i.e., where he measured $|1\rangle$ in the $z$-basis or $|-\rangle$ in the $x$-basis). Alice and Bob keep only the bits corresponding to these conclusive results.

A conclusive result occurs when Bob measures an orthogonal state, contributing a bit to the key. The probability of a conclusive measurement is given by:
\[
P(\text{conclusive}) = 1 - |\langle \psi_0 | \psi_1 \rangle|^2
\]
The overlap between the two quantum states is:
\[
\langle \psi_0 | \psi_1 \rangle = \frac{1}{\sqrt{2}}
\]
Thus, the probability of a conclusive measurement becomes:
\[
P(\text{conclusive}) = 1 - \left( \frac{1}{\sqrt{2}} \right)^2 = 1 - \frac{1}{2} = \frac{1}{2}
\]
On average, 50\% of the transmitted qubits contribute to the final key.

\textbf{Security check:} The security of the B92 relies on the no-cloning theorem \cite{wootters1982single} and the indistinguishability of non-orthogonal quantum states. An eavesdropper cannot perfectly distinguish between $|0\rangle$ and $|+\rangle$ states without disturbing them, which would introduce errors detectable by Alice and Bob.

\subsection{Combined Protocol}
Let $\rho AB$ be the final state shared between Alice and Bob after running the combined protocol. For $N$ rounds of the protocol, the final state can be described as: 
\begin{equation}
    \rho AB = \sum_{i = 1} ^ {N} (P_i GHZ_i + (1 - P_i) B92_i)
\end{equation}
Where, $P_i$ is the probability of selecting $GHZ$ protocol in round $i$, $GHZ_i$ is the state from $GHZ$ protocol in round $i$, $1 - P_i$ is the probability of selecting $B92$ protocol in round $i$ and $B92_i$ is the state from $B92$ protocol in round $i$.

\textbf{Fidelity Analysis: } Model for a noisy GHZ state in the presence of depolarizing noise is: 
\begin{equation}
    \rho GHZ = (1 - P) |GHZ\rangle \langle GHZ| + P\frac{I}{8}
\end{equation}

The fidelity between ideal and noisy GHZ state is $F_{GHZ}$:
\begin{equation}
    F(|GHZ\rangle, \rho GHZ) = \sqrt{\langle GHZ| \rho GHZ|GHZ \rangle}
\end{equation}

Model for noisy B92 state in the presence of depolarizing noise is:
\begin{align}
    \rho_0 = (1 - P) |\psi_0\rangle \langle \psi_0| + P\frac{I}{2}
    \\
    \rho_1 = (1 - P) |\psi_1\rangle \langle \psi_1| + P\frac{I}{2}
\end{align}

The fidelity between ideal and noisy B92 states are $F_{B92}$:
\begin{align}
     F(|\psi_0\rangle, \rho_0) = \sqrt{\langle \psi_0| \rho_0|\psi_0\rangle} \\
     F(|\psi_1\rangle, \rho_1) = \sqrt{\langle \psi_1| \rho_1|\psi_1\rangle}     
\end{align}

Combined protocol fidelity: 
\begin{equation}
    F_{combined} = min(F_{GHZ}, F_{B92})
\end{equation}

So, if $F_{combined}^2 > 1 - 2^{-s}$, then: $S(\rho) < (2n + s + \frac{1}{ln2})2^{-s} + O (2 ^ {-2s})$ \cite{nielsen2010quantum}, where $S(\rho)$ is the von Neumann entropy and $s$ is the security parameter.

\subsection{Protocol Abortion Conditions} 
The combined protocol must abort under any of the following conditions to ensure security:
\begin{itemize}
    \item $F_{combined}^2 \leq 1 - 2^{-s}$
    \item GHZ correlations deviate significantly where the equation \ref{ghzcorr1} and \ref{ghzcorr2} doesn't meet.
\end{itemize}
The protocol must immediately terminate if any of these conditions are met, ensuring that no compromised keys are used for encryption. These abortion conditions are necessary and sufficient to guarantee the security parameter for the final key.

\subsection{Key Generation Length of Combined QKD protocol}
\textbf{Entangled Mode (GHZ-based):} The probability of successful measurement in matching basis = $\frac{1}{8}$ (as all the tree parties must choose $\sigma_y$ basis). The raw key generation rate of Entangled mode is, $EM = \frac{n}{8}$. Then the final key length rate of entanlged mode is:
\begin{equation}
    EM_{final} = n (\frac{1}{8})(1 - h(\delta))
\end{equation}

\textbf{Non-Entangled Mode (B92-based):} The probability of successful measurement in matching basis = $\frac{1}{2}$ (matches basis selection between Alice and Bob). The raw key generation rate of Non-Entangled mode is, $NEM = \frac{n}{2}$. Then the final key length rate of non-entangled mode is:
\begin{equation}
    NEM_{final} = n (\frac{1}{2})(1 - h(\delta))
\end{equation}
Where, $h(\delta)$ is the binary entropy function and $\delta$ is the quantum bit error rate (QBER) \cite{elkouss2010information}.

\textbf{Combined Key Length:} Both entanled and non-entangled mode has $\frac{N}{2}$ rounds, where $N$ is the total number of $|C\rangle$ state. So, the expected total key length of the combined protocol is:

\begin{gather}
    CPL = P(C_1) \times EM_{final} + P(C_2) \times NEM_{final} \nonumber \\
    CPL = \frac{1}{2} \times (n (\frac{1}{8})(1 - h(\delta))) + \frac{1}{2} \times (n (\frac{1}{2})(1 - h(\delta))) \nonumber \\ 
    CPL = \frac{5n(1 - h(\delta)}{16}    
\end{gather}
Where $CPL$ is the total expected combined protocol key length, in equation \ref{c1prob} and \ref{c2prob} we have described about the probabilities of $C_1$ and $C_2$.

\section{Circuits Simulation Using Qiskit}
\subsection{Selection State}
The selection state is a crucial component of our combined protocol, determining which sub-protocol (GHZ or B92) will be used for each round of key generation.

Alice and Bob will measure the $|C\rangle$ state, which will choose them to use a random protocol between GHZ and B92 if the measurement results are same. 

\begin{figure}[!h]
    \centering
    \includegraphics[width=0.5\linewidth]{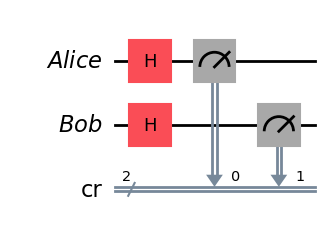}
    \caption{Selection state of combined protocol.}
    \label{fig:coin_toss}
\end{figure}

Figure: \ref{fig:coin_toss} shows the qiskit circuit diagram of selection state. Where we used a H(Hadamard)-gate to make the $|0\rangle$ state into a superposition state.  

\subsection{GHZ Protocol on Qiskit}
\begin{figure}[!h]
    \centering
    \includegraphics[width=0.4\linewidth]{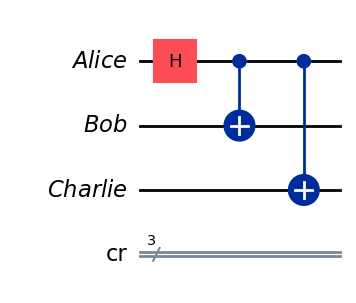}
    \caption{GHZ states.}
    \label{fig:ghz}
\end{figure}

If the selection states measurement gives $|C_1\rangle$ state for both Alice and Bob then they chooses to use GHZ QKD protocol. Figure: \ref{fig:ghz} shows the circuit diagram of GHZ states where we used H(Hadamard) and CNOT(Controlled Not) gates. 

\begin{figure}[!h]
\centering
    \begin{subfigure}{0.4\textwidth}
        \includegraphics[width=\textwidth]{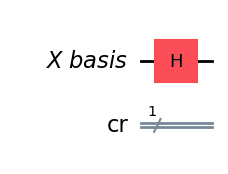}
        \caption{Pauli-$\sigma_x$ basis}
        \label{fig:x_basis}
    \end{subfigure}
    \hfill
    \begin{subfigure}{0.45\textwidth}
        \includegraphics[width=\linewidth]{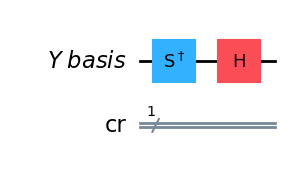}
        \caption{Pauli-$\sigma_y$ basis}
        \label{fig:y_basis}
    \end{subfigure}
\caption{Pauli basis.}
\label{fig:pauli_basis}
\end{figure}

\begin{figure}[!h]
    \centering
    \includegraphics[width=.9\linewidth]{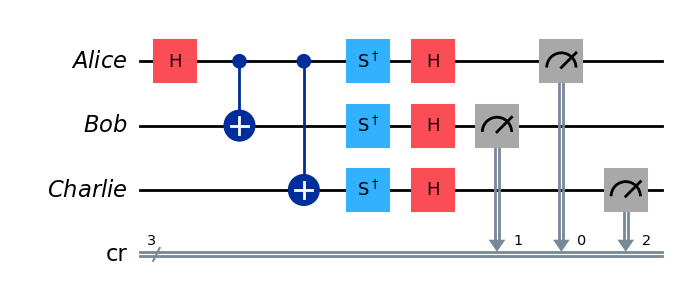}
    \caption{Alice, Bob and Charlie chooses $\sigma_y$ measurement basis to generate key.}
    \label{fig:measure_bit}
\end{figure}

\begin{figure}[!h]
\centering
    \begin{subfigure}{0.49\textwidth}
        \includegraphics[width=\textwidth]{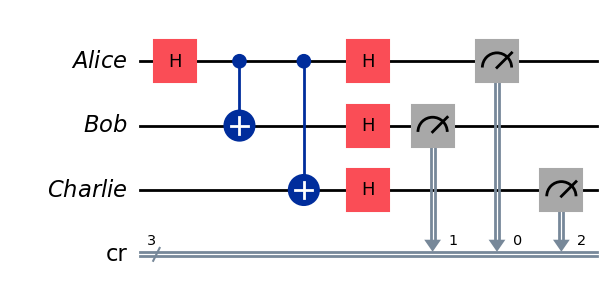}
        \caption{$X_AX_BX_C$ measurements}
        \label{fig:xxx}
    \end{subfigure}
    \hfill
    \begin{subfigure}{0.49\textwidth}
        \includegraphics[width=\linewidth]{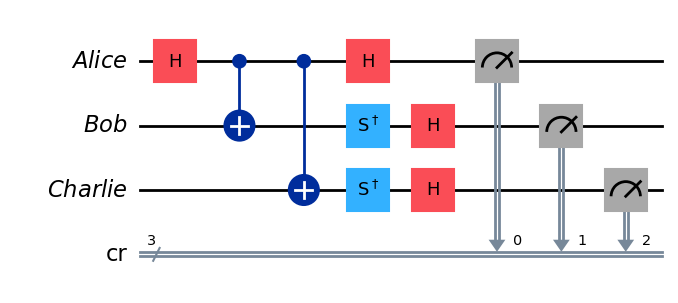}
        \caption{$X_AY_BY_C$ measurements}
        \label{fig:xyy}
    \end{subfigure}
    \begin{subfigure}{0.49\textwidth}
        \includegraphics[width=\textwidth]{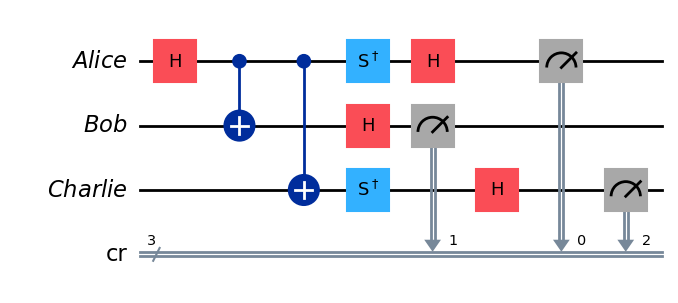}
        \caption{$Y_AX_BY_C$ measurements}
        \label{fig:yxy}
    \end{subfigure}
    \hfill
    \begin{subfigure}{0.49\textwidth}
        \includegraphics[width=\linewidth]{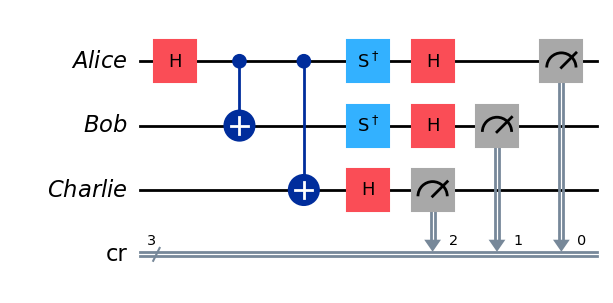}
        \caption{$Y_AY_BX_C$ measurements}
        \label{fig:yyx}
    \end{subfigure}
\caption{Correlation checking circuits for GHZ QKD.}
\label{fig:correlation_check}
\end{figure}

The GHZ-based QKD protocol employs a series of Pauli basis measurements to secure key generation and detect potential eavesdropping. Figure: \ref{fig:pauli_basis} presents the construction of the pauli-$\sigma_x$ and pauli-$\sigma_y$ basis, essential for measuring qubit states with high precision. Alice, Bob, and Charlie use these basis (with the $\sigma_y$ basis generated using $S^\dagger$ and Hadamard gates) for their respective qubits, ensuring robust correlation across all participants. Building on this, Figure: \ref{fig:measure_bit} illustrates the key generation process, where Alice, Bob, and Charlie consistently select the $\sigma_y$ basis for measurement. This alignment guarantees that the product of their outcomes remains -1, contributing securely to the key. In Figure: \ref{fig:correlation_check}, the focus shifts to the verification of these correlations through various basis combinations ($\sigma_x$, $\sigma_y$), serving as a security measure against eavesdroppers. The correlation checks reinforce the integrity of the GHZ states.

\subsection{B92 Protocol on Qiskit}
If the selection states measurement gives $|C_2\rangle$ state for both Alice and Bob then they chooses to use B92 QKD protocol.

\begin{figure}[!h]
\centering
    \begin{subfigure}{0.3\linewidth}
        \includegraphics[width=\linewidth]{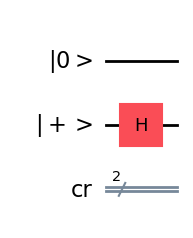}
        \caption{Alice prepares a state between $|0\rangle$ and $|+\rangle$ to send to Bob}
        \label{fig:0+state}
    \end{subfigure}
    \hfill
    \begin{subfigure}{0.52\linewidth}
        \includegraphics[width=\linewidth]{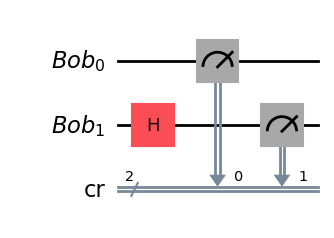}
        \caption{Bob's measurement in $\sigma_z$ and $\sigma_x$ basis}
        \label{fig:bobmeasureZX}
    \end{subfigure}
    \caption{Preparation of B92 circuits.}
    \label{fig:b92}
\end{figure}

We have implemented B92 protocol using Qiskit SDK, Figure: \ref{fig:b92} shows the circuit diagram of Alice's preparation of non-orthogonal states based on random bit string. And shows the circuit diagram of Bob's possible measurement in $\sigma_z$-basis or in $\sigma_x$-basis.

\section{Results and Analysis}
Our results and analysis is based on Qiskit SDK for Quantum Computing by IBM. We have compared the number of keys generated with different QKD protocol with our Combined QKD Protocol. We have also generated the numbers of getting $|C_1\rangle$ state and $|C_2\rangle$ state. Finally, We analysed the security measurement of our combined protocol.

\subsection{Key Analysis}
Here we have shown different measurement of key generation of different QKD protocol and compared those with our QKD protocol. And shown the measurement of $|C\rangle$ state.
Table: \ref{tab:numberofstate} shows the number of getting $|C_1\rangle$ and $|C_2\rangle$ states by measuring the $|C\rangle$ state. 
\begin{table}[!h]
    \centering
    \begin{tabular}{c|c|c}
        Number of measurement of $|C\rangle$ states & Number of $|C_1\rangle$ states (GHZ QKD) & Number of $|C_2\rangle$ states (B92 QKD) \\
        \hline
        5 & 2 & 3\\
        \hline
        10 & 5 & 5\\
        \hline
        15 & 8 & 7\\
        \hline
        20 & 10 & 10\\
    \end{tabular}
    \caption{Measuring $|C\rangle$ state.}
    \label{tab:numberofstate}
\end{table}

\begin{figure}[!h]
    \centering
    \includegraphics[width=0.8\linewidth]{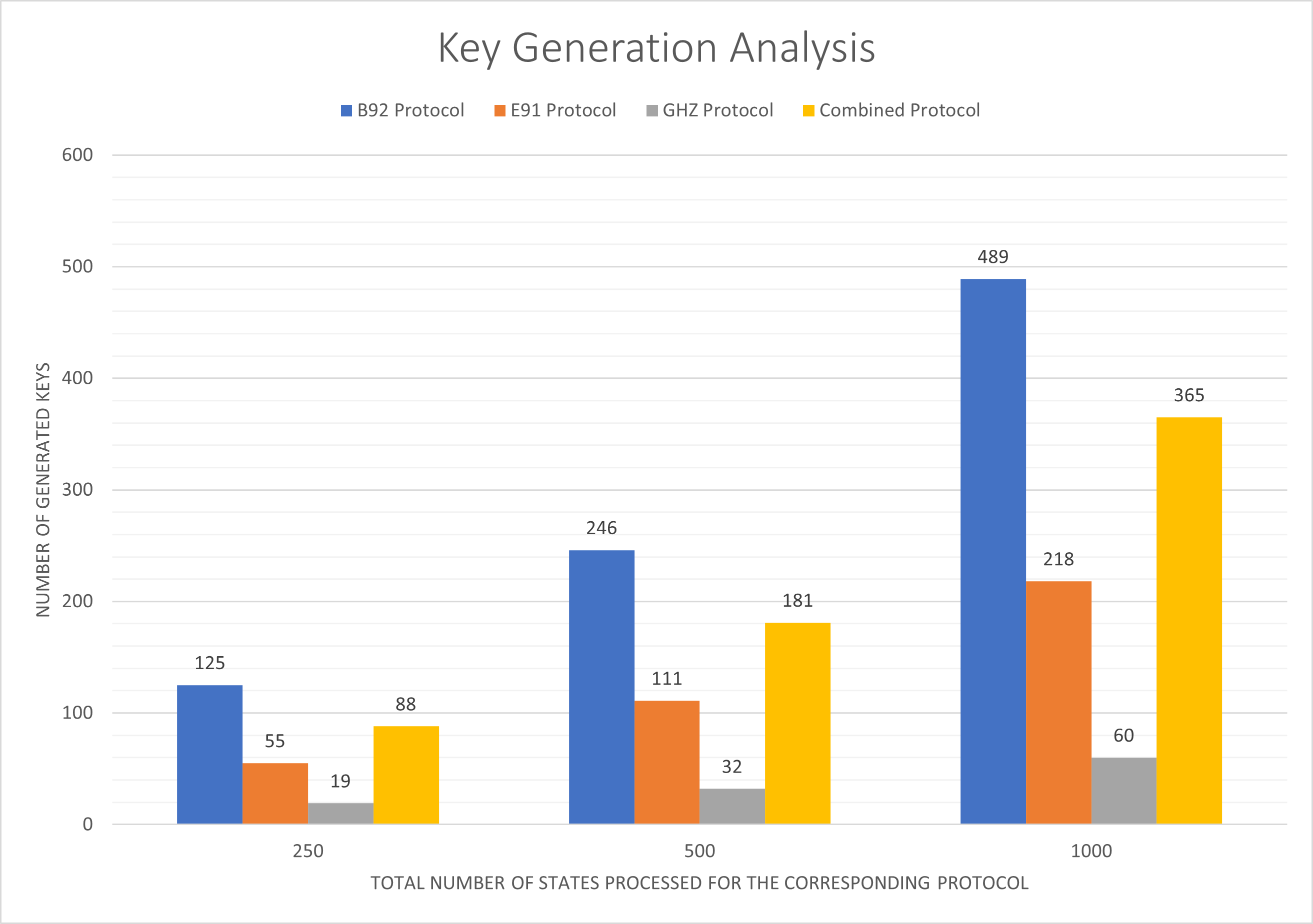}
    \caption{Comparison of number of key generation for different QKD}
    \label{fig:bit_gen}
\end{figure}

We have compared key generation analysis of different QKD protocol. Figure: \ref{fig:bit_gen} shows the comparison between different QKD protocol like B92, E91, GHZ and our Combined Protocol. Here we can see the number of key generation of our protocol is higher than entangled-based protocol like E91 and GHZ QKD but Non-entangled based protocol generates most number of keys. 

\begin{figure}[!h]
    \centering
    \includegraphics[width=0.8\linewidth]{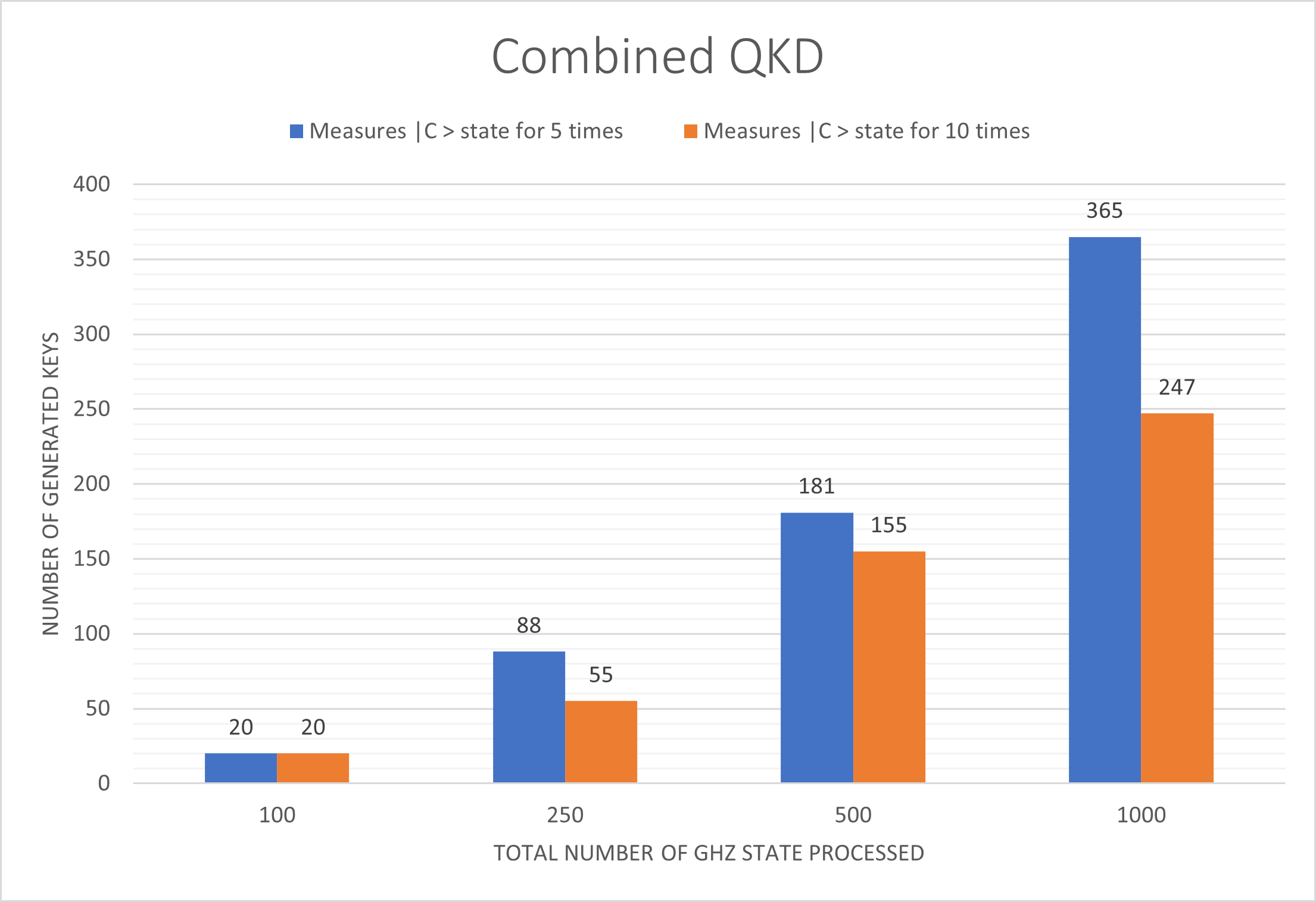}
    \caption{The number of key generated rates by number of states used for measuring $|C\rangle$ state}
    \label{fig:combined_qkd}
\end{figure}

We have also compared our Combined QKD protocol with it's different iteration/measurement of $|C\rangle$ state. Figure: \ref{fig:combined_qkd} shows the number of key generated rates by number of state processed for measuring $|C\rangle$ state for 5 and 10 times. The results suggest using 5 iteration of $|C\rangle$ state generates more numbers of key comparing to 10 iteration.

\subsection{Security Analysis}
B92 doesn't use entanglement. So, it's security comes from non-orthogonal random states \cite{elboukharireview2010}. Our Combined QKD combines both B92 and GHZ QKD, hence it allows to check security through strong quantum correlation.

\textbf{Maximal Entanglement:} The three-qubit GHZ state is defined as:

\begin{equation}
    |GHZ\rangle = \frac{1}{\sqrt{2}}(|000\rangle + |111\rangle)
\end{equation}

The density matrix of this state is given by:

\begin{equation}
    \rho_{\text{GHZ}} = |GHZ\rangle \langle GHZ|
\end{equation}

The von Neumann entropy \cite{zachos2007classical} for any bipartition is maximized:

\begin{equation}
    S(\rho_A) = S(\rho_B) = S(\rho_C) = 1
\end{equation}

where the von Neumann entropy \(S(\rho)\) is defined as:

\begin{equation}
    S(\rho) = - \text{Tr}(\rho \log_2 \rho)
\end{equation}

\textbf{Violation of Local Realism:} Consider the following observables for the three qubits:

\[
    A = \sigma_x, \quad B = \sigma_y, \quad C = \sigma_y
\]

Quantum mechanics predicts:

\begin{equation}
    \langle GHZ| ABC |GHZ\rangle = -1
\end{equation}

However, any local hidden variable theory would predict:

\begin{equation}
    (A_1 B_1 C_1)(A_1 B_2 C_2)(A_2 B_1 C_2)(A_2 B_2 C_1) = 1
\end{equation}

where \(A_1, A_2, B_1, B_2, C_1, C_2 = \pm 1\) are predetermined values. The contradiction between the quantum mechanical prediction and the local hidden variable theory demonstrates the violation of local realism.

\section{Conclusion}
This paper has presented a novel hybrid quantum key distribution protocol that combines the strengths of entanglement-based (GHZ state) and non-entanglement based (B92) approaches. Our research aimed to optimize the trade-off between security and key generation rate in QKD systems. We successfully designed and implemented a protocol that dynamically switches between GHZ-state and B92 methods using a quantum superposition state. This approach allows for adaptive QKD that can potentially adjust to varying channel conditions and security requirements.

Our combined protocol demonstrated higher key generation rates compared to standalone entanglement-based protocols like E91 and GHZ QKD. While it did not surpass the key generation rate of B92 alone, it offered a better balance between rate and security. By incorporating the GHZ-state component, our protocol leverages the strong quantum correlations and violation of local realism inherent in three-particle entanglement. This provides a robust security check mechanism not available in non-entanglement based protocols like B92.
We successfully simulated our protocol using Qiskit, demonstrating its feasibility on current quantum computing platforms. This implementation provides a foundation for further experimental work and potential real-world applications. Our results suggest that the protocol's performance can be tuned by adjusting the frequency of switching between GHZ and B92 components. This flexibility allows for optimization based on specific use-case requirements. The protocol efficiently utilizes both entangled and non-entangled quantum states, potentially making it more resource-effective than protocols relying solely on entanglement.

While our work shows promising results, further research is needed to fully characterize the protocol's performance under various noise models and against specific attack strategies. Additionally, experimental implementation beyond simulation will be crucial for validating the protocol's practical viability.

In conclusion, our hybrid QKD protocol represents a significant step towards more flexible and efficient quantum key distribution systems. By combining entanglement-based and non-entanglement based methods, we have demonstrated a promising approach to balancing security and efficiency in quantum cryptography. This work contributes to the ongoing effort to make QKD more practical and adaptable, potentially accelerating the development and deployment of large-scale quantum-secure communication networks.

\section*{Data availability statement}
Data will be made available on request.

\section*{Acknowledgement}
The authors would like to thank Mr. Md Shawmoon Azad, Research Assistant, North South
University, Mr. Sowmitra Das, Senior Lecturer, Department of CSE, BRAC University, and Syed
Emad Uddin Shubha, Teaching Assistant, BRAC University  for their valuable suggestions and constant support.

\section*{Competing Interests}
The author declares that there are no competing financial interests.

%Bibliography
\bibliographystyle{unsrt}  
\bibliography{references}

\begin{thebibliography}{10}

\bibitem{Tomamichel2017largelyself}
Marco Tomamichel and Anthony Leverrier.
\newblock A largely self-contained and complete security proof for quantum key distribution.
\newblock {\em {Quantum}}, 1:14, July 2017.

\bibitem{ekert1991quantum}
Artur~K Ekert.
\newblock Quantum cryptography based on bell’s theorem.
\newblock {\em Physical review letters}, 67(6):661, 1991.

\bibitem{bennett2014quantum}
Charles~H Bennett and Gilles Brassard.
\newblock Quantum cryptography: Public key distribution and coin tossing.
\newblock {\em Theoretical computer science}, 560:7--11, 2014.

\bibitem{bennett1992quantum}
Charles~H Bennett.
\newblock Quantum cryptography using any two nonorthogonal states.
\newblock {\em Physical review letters}, 68(21):3121, 1992.

\bibitem{greenberger1989going}
Daniel~M Greenberger, Michael~A Horne, and Anton Zeilinger.
\newblock Going beyond bell’s theorem.
\newblock In {\em Bell’s theorem, quantum theory and conceptions of the universe}, pages 69--72. Springer, 1989.

\bibitem{QiskitCommunity2017}
{Qiskit Community}.
\newblock Qiskit: {{An}} open-source framework for quantum computing, March 2017.

\bibitem{scarani2009security}
Valerio Scarani, Helle Bechmann-Pasquinucci, Nicolas~J Cerf, Miloslav Du{\v{s}}ek, Norbert L{\"u}tkenhaus, and Momtchil Peev.
\newblock The security of practical quantum key distribution.
\newblock {\em Reviews of modern physics}, 81(3):1301--1350, 2009.

\bibitem{einstein1935can}
Albert Einstein, Boris Podolsky, and Nathan Rosen.
\newblock Can quantum-mechanical description of physical reality be considered complete?
\newblock {\em Physical review}, 47(10):777, 1935.

\bibitem{bennett1984proceedings}
Charles~H Bennett, Gilles Brassard, et~al.
\newblock Proceedings of the ieee international conference on computers, systems and signal processing, 1984.

\bibitem{phillips1992heisenberg}
Melba Phillips.
\newblock Heisenberg and uncertainity.
\newblock {\em The Physics Teacher}, 30(1):39--40, 1992.

\bibitem{koashi1998no}
Masato Koashi and Nobuyuki Imoto.
\newblock No-cloning theorem of entangled states.
\newblock {\em Physical review letters}, 81(19):4264, 1998.

\bibitem{Horodecki_2009}
Ryszard Horodecki, Paweł Horodecki, Michał Horodecki, and Karol Horodecki.
\newblock Quantum entanglement.
\newblock {\em Reviews of Modern Physics}, 81(2):865–942, June 2009.

\bibitem{PhysRevLett.92.057901}
Valerio Scarani, Antonio Ac\'{\i}n, Gr\'egoire Ribordy, and Nicolas Gisin.
\newblock Quantum cryptography protocols robust against photon number splitting attacks for weak laser pulse implementations.
\newblock {\em Phys. Rev. Lett.}, 92:057901, Feb 2004.

\bibitem{bruss1998optimal}
Dagmar Bru{\ss}.
\newblock Optimal eavesdropping in quantum cryptography with six states.
\newblock {\em Physical Review Letters}, 81(14):3018, 1998.

\bibitem{von2013mathematische}
John Von~Neumann.
\newblock {\em Mathematische grundlagen der quantenmechanik}, volume~38.
\newblock Springer-Verlag, 2013.

\bibitem{zukowski1998quest}
Marek {\.Z}ukowski, A~Zeilinger, MA~Horne, and H~Weinfurter.
\newblock Quest for ghz states.
\newblock {\em Acta Physica Polonica A}, 93(1):187--195, 1998.

\bibitem{guo2010secure}
Ying Guo, Ronghua Shi, and Guihua Zeng.
\newblock Secure networking quantum key distribution schemes with greenberger--horne--zeilinger states.
\newblock {\em Physica Scripta}, 81(4):045006, 2010.

\bibitem{bell1964physics}
JS~Bell.
\newblock Physics long island city.
\newblock {\em NY}, 1(195):1, 1964.

\bibitem{storz2023loophole}
Simon Storz, Josua Sch{\"a}r, Anatoly Kulikov, Paul Magnard, Philipp Kurpiers, Janis L{\"u}tolf, Theo Walter, Adrian Copetudo, Kevin Reuer, Abdulkadir Akin, et~al.
\newblock Loophole-free bell inequality violation with superconducting circuits.
\newblock {\em Nature}, 617(7960):265--270, 2023.

\bibitem{cafaro2024violationbellsinequalityclauserhorneshimonyholt}
Carlo Cafaro, Christian Corda, Philip Cairns, and Ayhan Bingolbali.
\newblock Violation of bell's inequality in the clauser-horne-shimony-holt form with entangled quantum states revisited, 2024.

\bibitem{bennett1992quantumbbm}
Charles~H Bennett, Gilles Brassard, and N~David Mermin.
\newblock Quantum cryptography without bell’s theorem.
\newblock {\em Physical review letters}, 68(5):557, 1992.

\bibitem{mcmahon2008quantum}
David McMahon.
\newblock {\em Quantum computing explained}.
\newblock John Wiley \& Sons, 2008.

\bibitem{molina2005experimental}
G~Molina-Terriza, A~Vaziri, R~Ursin, and A~Zeilinger.
\newblock Experimental quantum coin tossing.
\newblock {\em Physical review letters}, 94(4):040501, 2005.

\bibitem{greenberger1990bell}
Daniel~M Greenberger, Michael~A Horne, Abner Shimony, and Anton Zeilinger.
\newblock Bell’s theorem without inequalities.
\newblock {\em American Journal of Physics}, 58(12):1131--1143, 1990.

\bibitem{liu2021experimental}
Zheng-Hao Liu, Jie Zhou, Hui-Xian Meng, Mu~Yang, Qiang Li, Yu~Meng, Hong-Yi Su, Jing-Ling Chen, Kai Sun, Jin-Shi Xu, et~al.
\newblock Experimental test of the greenberger--horne--zeilinger-type paradoxes in and beyond graph states.
\newblock {\em npj Quantum Information}, 7(1):66, 2021.

\bibitem{wootters1982single}
William~K Wootters and Wojciech~H Zurek.
\newblock A single quantum cannot be cloned.
\newblock {\em Nature}, 299(5886):802--803, 1982.

\bibitem{nielsen2010quantum}
Michael~A Nielsen and Isaac~L Chuang.
\newblock {\em Quantum computation and quantum information}.
\newblock Cambridge university press, 2010.

\bibitem{elkouss2010information}
David Elkouss, Jesus Martinez-Mateo, and Vicente Martin.
\newblock Information reconciliation for quantum key distribution.
\newblock {\em arXiv preprint arXiv:1007.1616}, 2010.

\bibitem{elboukharireview2010}
Mohamed Elboukhari, Mostafa Azizi, and Abdelmalek Azizi.
\newblock Quantum key distribution protocols: A survey.
\newblock 01 2010.

\bibitem{zachos2007classical}
Cosmas~K Zachos.
\newblock A classical bound on quantum entropy.
\newblock {\em Journal of Physics A: Mathematical and Theoretical}, 40(21):F407, 2007.

\end{thebibliography}

\end{document}